\documentclass[12pt]{article}

\usepackage[T1]{fontenc}
\usepackage{tgtermes}
\usepackage{gensymb}
\usepackage{titling}
\usepackage{authblk}

\usepackage{graphicx}
\usepackage{amssymb}
\usepackage{bbm}
\usepackage{amsmath}
\usepackage{mathtools}
\usepackage{cancel}
\usepackage{slashed}
\usepackage[colorlinks,citecolor=blue,urlcolor=blue, linkcolor=blue, bookmarks=false,hypertexnames=true]{hyperref}

\usepackage{cite}
\usepackage{hyperref}
\usepackage{sidecap}
\usepackage{caption}
\usepackage{subcaption}
\usepackage{geometry}

\title{Hausdorff dimension of fermions on a random lattice}

\author[a]{Mattia Varrone \thanks{\href{mailto:mgv29@cam.ac.uk}{mgv29@cam.ac.uk}}}
\author[b,c]{William E. V. Barker\thanks{\href{mailto:wb263@cam.ac.uk}{wb263@cam.ac.uk}}}
\affil[a]{Department of Applied Mathematics and Theoretical Physics, University of Cambridge}
\affil[b]{Astrophysics Group, Cavendish Laboratory, JJ Thomson Avenue, Cambridge CB3 0HE, UK}
\affil[c]{Kavli Institute for Cosmology, Madingley Road, Cambridge CB3 0HA, UK}
\predate{}
\postdate{}
\date{}

\begin{document}
    \pagenumbering{arabic}
    	\maketitle

    	\begin{abstract}
        \noindent
        Geometric properties of lattice quantum gravity in two dimensions are studied numerically via Monte Carlo on Euclidean Dynamical Triangulations. A new computational method is proposed to simulate gravity coupled with fermions, which allows the study of interacting theories on a lattice, such as non-Riemannian gravity models. This was tested on Majorana spinors, where we obtained a Hausdorff dimension $d_{W} = 4.22 \pm 0.03$, consistent with the bounds from the literature $ {4.19 < d_{H} < 4.21}$.
    	\end{abstract}

    
	\section{Introduction}
    Lattices are a natural setting to study strongly interacting quantum field theories, and gravity is no exception. Indeed, discrete systems with finite size have a well-defined path integral, and in theories with Euclidean signature, every field configuration is associated with a Boltzmann factor carrying a probability interpretation.
    In this paper, we discretize 2D Euclidean gravity on spherical topology coupled with fermions and we generate an ensemble of possible geometries via dynamical triangulation, using Markov Chain Monte Carlo (MCMC) techniques, as prescribed in \cite{buddMonte, EDT}. 
    The resulting manifold, called a \textit{simplicial manifold}, reproduces the critical exponents of Liouville quantum gravity when the system is coupled with conformal matter \cite{liouville}.
    The methods used here will be founded in the triangulated fermion construction of Burda, Bogacz, Jurkiewicz, Krzywicki, Petersen and Petersson (BBJKPP) in their seminal work ~\cite{burda1999wilson,burda1999fermions,bogacz2001dirac,bogacz2002spectrum,bogacz2003fermions,bilke19984d,Burda:1989mk}.
    We will use a similar constructions to simulate free Majorana spinors and measure their effect on the Hausdorff dimension of the manifold. 
    In the past, this was always achieved by mapping the fermionic system to a proxy Ising model ~\cite{bogacz2001dirac,bogacz2002spectrum} (indeed, both Majorana spinors and critical the Ising model have central charge $c = \frac{1}{2}$). Instead, we decided to compute the spinorial action directly by calculating the determinant of the Dirac\textendash Wilson operator, as suggested in \cite{bogacz2002spectrum}.
    This has the advantage of allowing for more general interactions between gravity and matter, such as through non-Riemannian gauge fields or higher-point fermionic vertices.
    Using this method, we compute a value of the Hausdorff dimension $d_{W} = 4.22 \pm 0.03$ which differs from the previous value computed by Bogacz and Burda\ in \cite{bogacz2002spectrum} $d_{W} = 2.87$, while favouring the conjectural relations between the Hausdorff dimension and central charge proposed by Watabiki in \cite{watabiki1993analytic}, yielding $d_{W} \approx 4.2122$, and it is consistent with the bounds rigorously derived by Gwynne $4.1892 < d_{W} < 4.2156$\ \cite{gwynne2019bounds}.\\
    The paper content is organized as follows: firstly, we introduce the essential ingredients to represent fermions on EDT. Secondly, we discuss how Majorana spinors affect the path integral and give different weights to triangulations. 
    Then our measurements for the Hausdorff dimension are presented and compared to the literature. Finally, generalizations of the model are proposed.

    
\section{Local frames and spinor transport}\label{local frames}
\begin{figure}[h!]
    \begin{center}
    \includegraphics[width=3.5in]{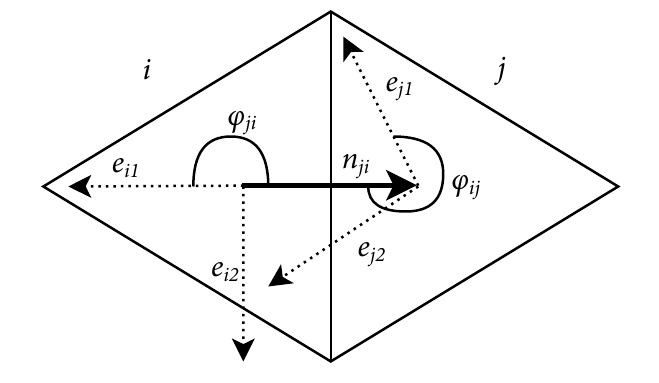}
    \caption{The local frames associated with two neighbouring triangles, $i$ and $j$, are represented by orthonormal basis vectors $e_{1}$ and $e_{2}$. The connecting vector $n_{ij}$ is also shown. Since the triangles are equilateral, we can deduce from the figure that: $\phi_{ji} = \pi$, $\phi_{ij} = \frac{5\pi}{3}$, and $\Delta\phi_{ij} = \frac{5\pi}{3}$}.
    \label{fig:local_frame}
    \end{center}
    \end{figure}

    To represent fermions on a lattice we require local frames. On a spherical triangulation, this is achieved simply by defining a right-handed basis of orthonormal vectors $e_{i1}$ and $e_{i2}$ at the centre of every triangle $i$. We will use the convention proposed in \cite{bogacz2001dirac}, as presented in figure \ref{fig:local_frame}.
    To complete the spin structure, we also need a way to map the frames of neighbouring triangles: a parallel transport.
    For triangles $i$ and $j$, we define the unit vector $n_{ji}$ pointing in direction $i \to j$, and the angles $\phi_{ji}$ and $\phi_{ij}$, measured clockwise respectively from the frames $e_{i1}$ and $e_{j1}$ to $n_{ji}$.  
    For propaedeutic purposes, let us first consider the vector parallel transport  from triangle $i$ to $j$ with relative angle $\Delta \phi_{ij}$, which is generated by the element of the $\mathfrak{so}(2)$ Lie algebra $\epsilon$, the standard 2D antisymmetric tensor
        \begin{equation}\label{eq=transport_unflagged}
        U_{ij} \equiv e^{\epsilon \Delta \phi_{ij}} =\begin{pmatrix}
    \cos{\Delta \phi_{ij}} & \sin{\Delta \phi_{ij}} \\ \\
    -\sin{\Delta \phi_{ij}} & \cos{\Delta \phi_{ij}}
    \end{pmatrix}.
        \end{equation}
        
        Following Bogacz's prescription, we choose a representation of fermions such that the spinor parallel transport operator $i \to j$,  $\:\mathcal{U}_{ij}$, satisfying $\mathcal{U}_{ij}^{2} = U_{ij}$. Hence, we obtain
        \begin{equation}\label{eq=transport_spinor}
            \mathcal{U}_{ij} \equiv s_{ij} \: e^{\frac{\epsilon \Delta \phi_{ij}}{2}} = s_{ij} \begin{pmatrix}
        \cos{\frac{\Delta \phi_{ij}}{2}} & \sin{\frac{\Delta \phi_{ij}}{2}} \\ \\
        -\sin{\frac{\Delta \phi_{ij}}{2}} & \cos{\frac{\Delta \phi_{ij}}{2}}
    \end{pmatrix},
        \end{equation}
        where $\Delta \phi_{ij} \equiv \phi_{ij}-\phi_{ji}+\pi$ and the factors $s_{ij}$, taking values $+1$ or $-1$, are called \textit{sign flags}. Moreover, the parallel transport from $i \to j$ and then $j \to i$ leaves spinors unchanged, we must obtain the identity $\mathbbm{1}$, hence
        \begin{equation}
            \mathcal{U}_{ij} \mathcal{U}_{ji} = s_{ij}s_{ji} \: e^{\frac{ \Delta \phi_{ij} + \Delta \phi_{ji}}{2}\epsilon} = s_{ij}s_{ji} \: e^{\pi\epsilon} = -s_{ij}s_{ji} \mathbbm{1},
        \end{equation}
        implying
        \begin{equation}\label{eq=sign_inverse}
        s_{ij}s_{ji} = -1 \Leftrightarrow s_{ij} = - s_{ji}.
        \end{equation}
        Unsurprisingly, spinor parallel transport has the effect of rotating field components by half the amount in comparison to vector parallel transport.
    The ambiguity in the value of the sign flags is resolved by enforcing the physical consistency of spinor transport. Indeed, the trace of the parallel transport along an elementary counter-clockwise $n$-loop around a vertex $P$ over a set of triangles $\{i_{1}, i_{2}, ..., i_{n}, i_{1}\}$ (in lattice gauge theory, this is referred to as a \textit{plaquette}) can be related to the curvature at $P$, characterized by a \textit{deficit angle} $\Delta_{P}$. In particular, one obtains:
    \begin{equation}\label{eq=plaquette}
    \Pi_{P} \equiv \frac{1}{2} \text{Tr} \;  \mathcal{U}_{i_{n}i_{1}}... 
    \mathcal{U}_{i_{2}i_{3}} \mathcal{U}_{i_{1}i_{2}} = S_{P} \cos{\frac{\Delta_{P}}{2}}.
    \end{equation}
    We recall that the deficit angle is defined to be the difference between a full circle $2\pi$ and the total angle obtained by summing the angles having $P$ as a vertex, so for a $n$-plaquette on an equilateral triangulation we have $\Delta_{P} = 2\pi - \frac{n\pi}{3}$.
    The term $S_{P}$ can take values $+1$ or $-1$, depending on all elementary transports on the plaquette. On physical grounds, as explained in \cite{bogacz2001dirac}, we require the following for all elementary loops:
    \begin{equation}\label{eq=sign_loop}
     S_{P} = +1, \qquad \forall P.
    \end{equation}
    Conditions \eqref{eq=sign_inverse} and \eqref{eq=sign_loop} determine the sign flags uniquely, and they must be preserved throughout the evolution of the system. We will show in section \ref{flags} how to construct a suitable initial sign flag configuration and how to update it as the system evolves. \\
    This representation is compatible with the Majorana representation of the gamma matrices
    \begin{equation}\label{eq=gamma}
        \gamma_{1} =  \sigma_{3} \equiv \begin{pmatrix}
            1 & 0 \\
            0 & -1
        \end{pmatrix}, \qquad 
        \gamma_{2} =  \sigma_{1} \equiv \begin{pmatrix}
            0 & 1 \\
            1 & 0
        \end{pmatrix}.
    \end{equation}	

    It is useful to recall how the $c=1/2$ conformal field theory emerges from among the Majorana components $\Psi_\alpha$ in \eqref{eq=majorana}, for which we can write $\Psi^{\mathrm{T}}=(\Psi_1,\Psi_2)$. In the continuum, and with the real Clifford basis chosen in~\eqref{eq=gamma}, we hope to recover the free Majorana Lagrangian in Cartesian coordinates
\begin{equation}\label{free_action}
	L = \frac{1}{2}\bar{\Psi}\slashed{\partial}\Psi
	=\frac{1}{2}i\Psi^{\mathrm{T}}\sigma_2\left(\sigma_3\partial_x+\sigma_1\partial_y\right)\Psi.
\end{equation}
Using the complexified coordinates $z\equiv x+iy$ and $\bar{z}\equiv x-iy$ with $\partial_z\equiv \frac{1}{2}\left(\partial_x-i\partial_y\right)$ and $\partial_{\bar{z}}\equiv \frac{1}{2}\left(\partial_x+i\partial_y\right)$, we can obtain from~\eqref{free_action} the Lagrangian in the component form
\begin{equation}
	\begin{aligned}
		L=&\
	\frac{i}{2}\left(\Psi_1+i\Psi_2\right)\partial_z\left(\Psi_1+i\Psi_2\right)\\
		&\ +\frac{i}{2}\left(i\Psi_1+\Psi_2\right)\partial_{\bar{z}}\left(i\Psi_1+\Psi_2\right).
	\label{free_action_conformal}
	\end{aligned}
\end{equation}
The terms in \eqref{free_action_conformal} can be identified with a new pair of Grassmann numbers, related to the original variables by a transformation with unit determinant, which does not affect the functional measure in the path integral. In terms of these new variables the field equations become
\begin{equation}
	\partial_z\left(\Psi_1+i\Psi_2\right)/\sqrt{2}=0,\quad
	\partial_{\bar{z}}\left(i\Psi_1+\Psi_2\right)/\sqrt{2}=0,
	\label{field_equations}
\end{equation}
and these act to enforce antiholomorphic and holomorphic solutions.
    
    \subsection{General spin-connections}
    As a side note, we mention that our work is motivated partly by the aim of introducing more general gravitational connections on triangulations, particularly of non-Riemannian character.
    For this purpose, we note that in 2D the most general spin connection components  $\omega_{abc}$ in a local frame are given by \cite{blagojevic2dtorsion}
    \begin{equation}
        \omega_{abc} = \epsilon_{ab}\Omega_{c}.
    \end{equation}
    Moreover, if we allow the spin connection to be torsionful, we can decompose it into the Ricci rotation coefficients $\Gamma_{abc} \equiv \epsilon_{ab}A_{c}$ and the contorsion tensor $K_{abc} \equiv \epsilon_{ab}\mathcal{A}_{c}$ \cite{blagojevic2001gravitation}, as
    \begin{equation}
    \omega_{abc} = \Gamma_{abc} + K_{abc} = \epsilon_{ab} (A_{c} + \mathcal{A}_{c}).
    \end{equation}If we consider parallel transport of vector fields, we have a vector connection proportional to the 2D rotation generator $\epsilon$, given by \boldmath${\omega}$\unboldmath$_{c} = (A_{c} + \mathcal{A}_{c})\epsilon$. Therefore, the parallel transport operator on the triangulation between two triangles $i$ and $j$, in the direction $n_{ji}$, reads
    \begin{equation}
    U_{ij}^{(\omega)} = e^{\epsilon\int_{i}^{j} dx (n_{ji})^{c} (A_{c} + \mathcal{A}_{c})} = e^{(\Delta\phi_{ij} + \theta_{ij})\epsilon}
    .\end{equation}
    Accordingly, the Ricci rotation coefficients determine the rotation angle $\Delta\phi_{ij}$ for vector field components in the local frame basis, and the contorsion contributes an additional angle $\theta_{ij}$.
    For the transport of spinors, we obtain an analogous formula:
    \begin{equation}\label{eq=torsiontransport}
    \mathcal{U}_{ij}^{(\omega)} = s_{ij}\:e^{\frac{\Delta\phi_{ij} + \theta_{ij}}{2}\epsilon}= s_{ij} \begin{pmatrix}
        \cos{\frac{\Delta \phi_{ij} + \theta_{ij}}{2}} & \sin{\frac{\Delta \phi_{ij} + \theta_{ij}}{2}} \\ \\
        -\sin{\frac{\Delta \phi_{ij} + \theta_{ij}}{2}} & \cos{\frac{\Delta \phi_{ij} + \theta_{ij}}{2}}
    \end{pmatrix}.
    \end{equation}
    Therefore, the presence of "torsion" requires us to keep track of additional variables $\theta_{ij}$ linking neighbouring triangles. These introduce additional degrees of freedom in our description of physics, and imply that we have gauged the group $SO(2)$.
    \subsection{Fixing sign flags}\label{flags}
\begin{figure}[h!]
    \begin{center}
    \includegraphics[angle=0]{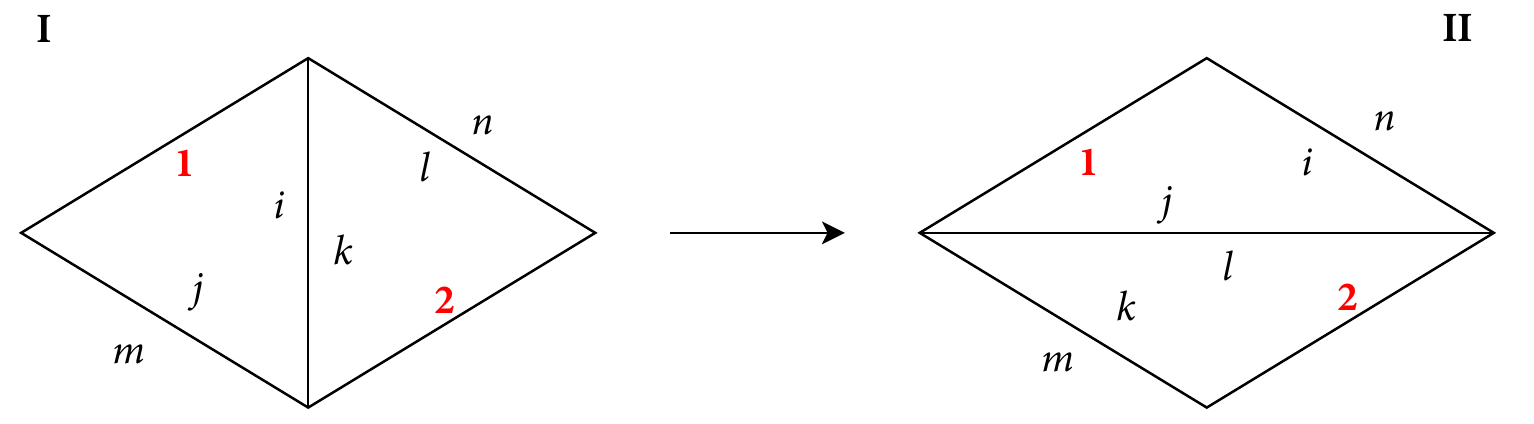}
    \caption{A flip move changes the connectivity of triangles from configuration \textbf{I} to \textbf{II}. Labels $1$ and $2$ identify the triangles before and after the transformation, while latin letters indicate their edges.}
    \label{fig:flip_move}
    \end{center}
    \end{figure}

\begin{figure}[h!]
    \begin{center}
    \includegraphics{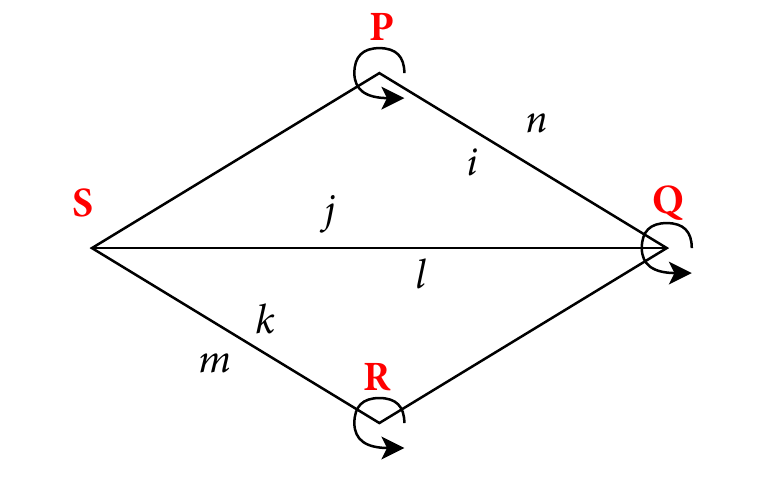}
    \caption{After a flip move is performed, we can restore the consistency condition of the flag configuration by calculating the plaquettes corresponding to vertices, $P, Q, R$ and sequentially updating the sign flags of edges $i, j, k$.}
    \label{fig:flag update}
    \end{center}
    \end{figure}

If we discretize euclidean 2D space as a spherical triangulation with a fixed number of triangles $N$, all possible geometries can be generated from MCMC with a finite number of \textit{flip moves} (also known as $(2, 2)$-moves) as described by Ambjørn \ in \cite{ambjornflipmove} and Budd in \cite{buddMonte}. The effect of such moves is shown schematically in figure \ref{fig:flip_move}. 
To simulate the effect of fermions on spacetime, the quantities encoding the spin structure also need to evolve consistently.
In the rest of this section, we will outline an algorithm to update sign flags $s_{ij}$  as flip moves are performed, while preserving the validity of conditions \eqref{eq=sign_inverse} and \eqref{eq=sign_loop}. In addition, we will construct a consistent initial configuration of triangles and sign flags, which will serve as a starting point for MCMC simulations.
For later convenience, we will label parallel transporters and sign flags by the edges. For example, referring to configuration \textbf{I} in figure \ref{fig:flip_move}, the sign flag associated with the link $1 \to 2$ will be relabelled using the index of the connecting edge in the starting triangle: $s_{12} \rightarrow s_{i}$. \\  
Now, assume that we start from a valid configuration of spin flags on a certain triangulation, and that we perform a flip move, affecting the connectivity of two triangles, labelled as in figure \ref{fig:flip_move}.
This is a local change, and therefore it only affects the validity of the consistency conditions for edges and vertices of triangles $1$ and $2$. 
We will proceed by first fixing condition \eqref{eq=sign_inverse}, with the following assignment of sign flags, in the specified order:\\
\begin{align}
\begin{split}
    s^{\text{new}}_{i} \to &\ -s^{\text{old}}_{n}, \\
    s^{\text{new}}_{k} \to &\ -s^{\text{old}}_{m}, \\
    s^{\text{new}}_{m} \to &\ -s^{\text{new}}_{k}, \\
    s^{\text{new}}_{l} \to &\ -s^{\text{old}}_{i}, \\
    s^{\text{new}}_{j} \to &\ -s^{\text{new}}_{l}. \\
\end{split}
\end{align}
Next, we must ensure that condition \eqref{eq=sign_loop} is met by all the modified plaquettes, which are those corresponding to the vertices $P, Q, R, S$ as shown in figure \ref{fig:flag update}.  
We devised the following algorithm: starting from vertex P, we compute the trace of the plaquette $\Pi_{P}$ and the deficit angle $\Delta_{P}$. From these, the sign $S_{P}$ can be deduced using equation \eqref{eq=plaquette}.
Now, since the parallel transporter $\mathcal{U}_{i}$ contributes to $\Pi_{P}$, we can update the sign flags
\begin{align}
\begin{split}
    s_{i} \to s_{i} \cdot S_{P}\\
    s_{n} \to -s_{i},
\end{split}
\end{align}
where the second equation ensures that condition \eqref{eq=sign_inverse} remains true.

We can repeat these steps for vertices $Q$ and $R$, and modify $s_{j}, s_{k}$ accordingly, since these flags do not affect the plaquettes previously fixed.
Finally, a combinatoric argument layed out from Burda in \cite{burda1999wilson} guarantees that if all but one elementary loop in a triangulation of a sphere are known to satisfy conditions \eqref{eq=sign_inverse} and \eqref{eq=sign_loop}, then also the last loop satisfies them, hence we are done.

\begin{figure}[h!]
    \begin{center}
    \includegraphics[width=5.5in]{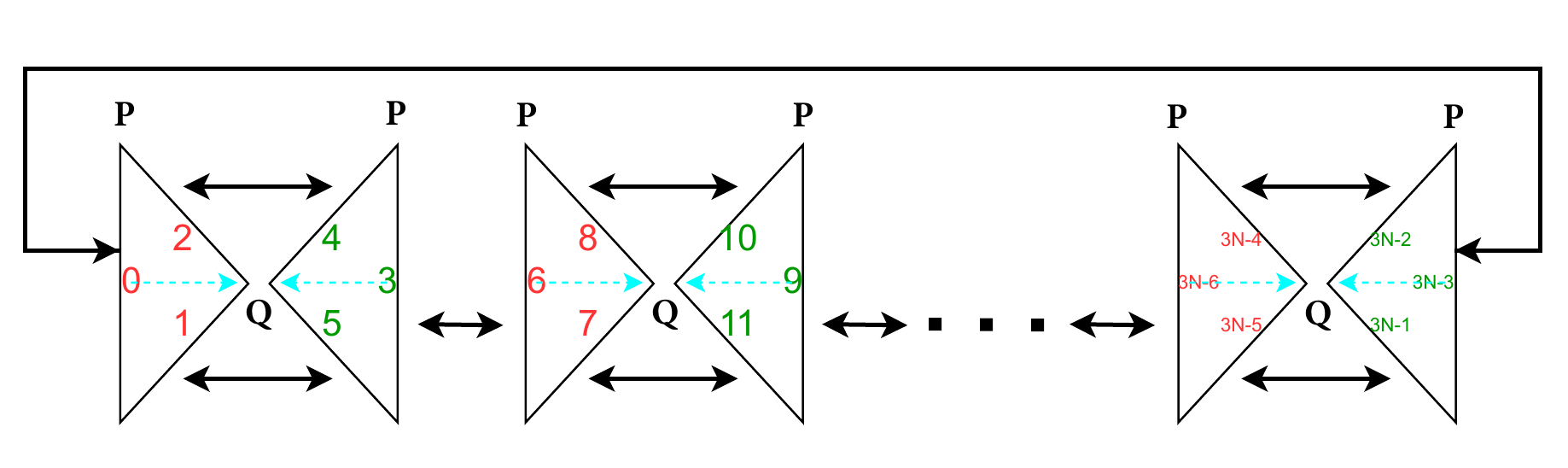}
    \caption{A fan triangulation composed of $N$ triangles is shown (the edge indexing convention can be found in \cite{buddMonte}). In green, we denoted edges carrying a positive sign flag $s = +1$, while flags $s = -1$ correspond to red edges. The direction of frame vectors $e_{1}$ is shown by cyan lines. This unambiguously identifies the spin structure of our triangulation. Labels $P$ and $Q$ denote the polar and equatorial vertices respectively.}
    \label{fig:fan triangulation}
    \end{center}
    \end{figure}

Now we are only left to show that a suitable triangulation with a consistent sign flag configuration exists.
Let us consider the explicit construction given by the \textit{fan triangulation} shown in figure \ref{fig:fan triangulation}.

    Here, we labeled with green the edges carrying a positive sign flag $s = +1$, while flags $s = -1$ correspond to red edges. The direction of frame vectors $e_{1}$ is shown by cyan lines. This unambiguously identifies the spin structure of our triangulation. 
 We will prove that the fan satisfies the consistency conditions whenever we choose a triangulation of size $N$, with $N \equiv 2 \;(\text{mod} 4)$.
We begin by noticing that we can classify vertices into only two types: \textit{equatorial vertices}, denoted by $Q$, are shared by two triangles, and two \textit{polar vertices}, denoted by $P$, shared by all $N$ triangles.
By the translation symmetry of the fan triangulation, we deduce that plaquettes associated with vertices Q share the same sign $S_{Q}$.
South and north pole are also equivalent due to mirror symmetry.\\
Let us calculate $S_{Q}$ for the vertex connected to the triangle with edges $0,1,2$. From formula \eqref{eq=plaquette} and the property of rotation matrices we know the plaquette's trace is given by $\Pi_{Q} = s_{1}s_{4}\cos{\frac{\Delta\phi_{1}+\Delta\phi_{4}}{2}} = - \cos{\frac{\Delta\phi_{1}+\Delta\phi_{4}}{2}}$. The spinor transport from edge $1$ to $5$ gives is associated to $\frac{\Delta\phi_{1}}{2} = \frac{1}{2}(\frac{5\pi}{3} - \frac{\pi}{3} + \pi) = \frac{7\pi}{6}$, and transport from edge $4$ to $2$ similarly gives $\frac{\Delta\phi_{1}}{2} = \frac{7\pi}{6}$, hence 
$\Pi_{Q}=-\cos(\frac{7\pi}{3})=-\frac{1}{2}$. On the other hand, the angle deficit is $\Delta_{Q} = \frac{4\pi}{3}$, thus $\cos{\frac{\Delta_{Q}}{2}} = -\frac{1}{2} = \Pi_{Q}$. Therefore, $S_{Q} = +1$. \\
Let us now check the same for the polar vertex. Due to the repeating pattern of the plaquette, we only need to compute: $\frac{\Delta\phi_{2}}{2}=\frac{1}{2}(\frac{\pi}{3} -\frac{5\pi}{3} + \pi) = -\frac{\pi}{6}$ and $\frac{\Delta\phi_{3}}{2}=\frac{1}{2}(\pi - \pi +\pi) = \frac{\pi}{2}$. Since there are $\frac{N}{2}$ tuples of triangles contributing the same parallel transport, we get $\Pi_{P} = (-1)^{\frac{N}{2}}\cos{\frac{N}{2}(\frac{\Delta\phi_{2}+\Delta\phi_{3}}{2})}=-(-1)^{\frac{N}{2}}\cos{\frac{N\pi}{6}} $. The deficit angle can be calculated easily as $\Delta_{P}=2\pi-\frac{N\pi}{3}$, thus $\cos{\frac{\Delta_{P}}{2}}=-\cos{\frac{N\pi}{6}}$. Finally, these quantities satisfy the consistency relations if $N \equiv 2 \;(\text{mod} 4)$, and this concludes the proof.

    
\section{Majorana spinors and path integral}\label{spinors}
    Let us now discuss how Majorana spinors are coupled with gravity on the lattice.
    On a given triangulation $T$, fermionic fields are characterized by degrees of freedom $\Psi_{i}$ and $\bar{\Psi}_{i}$ sitting at the centres of triangles $i$.
    A Majorana spinor $\Psi$ carries Grassmann components $\Psi_{\alpha}$, denoted by a spinor index $\alpha = 1, 2$ while the components of its conjugate spinor $\bar{\Psi}$ are represented with an upper index $\Psi^{\beta}$ and they are related via the Majorana condition
    \begin{equation}\label{eq=majorana}
    \Psi^{\beta} = \epsilon^{\beta\alpha}\Psi_{\alpha} \;\leftrightarrow\; \Psi_{\alpha} = \Psi^{\beta}\epsilon_{\beta\alpha}.
    \end{equation}
    The condition above implies that $\bar{\Psi}$ belongs to the dual space with respect to $\Psi$, and the antisymmetric tensor relates the two via charge conjugation.
    With the definition of parallel transport \eqref{eq=transport_spinor} and our choices of representation \eqref{eq=gamma} and \eqref{eq=majorana}, we have now all the ingredients to build an action for free fermions coupled to a triangulation $T$. This reads
        \begin{equation}\label{eq=action}
        	S_{T} = -K \sum_{\langle ij \rangle}\bar{\Psi}_{i} H_{ij} \Psi_{j} + \frac{1}{2}\sum_{i}\bar{\Psi}_{i}  \Psi_{i}.
        \end{equation}
    The first contribution is a regularised kinetic term, called \textit{hopping term}, with $K$ a constant and $H_{ij} = \frac{1}{2}(1 + n_{ij}^{a} \: \gamma_{a})\:\mathcal{U}_{ij}$.
    The second mass term is a counter-term introduced in order to cancel the effective mass induced by the presence of a lattice, and tune the systems to criticality.
    Conformal symmetry is essential to reproduce the behaviour expected from Liouville theory, and this is achieved when $K \approx 0.3746$ \cite{bogacz2002spectrum}.
    We can rewrite the action in a more compact form by employing the \textit{Dirac-Wilson operator} $D_{ij}$. Making spinor indices explicit
    \begin{align}\label{eq=diracwilson}
    \begin{split}
     S_{T} = \sum_{i, j}\bar{\Psi}_{i} D_{ij} \Psi_{j} = \sum_{i, j}{\Psi}_{i}^{\alpha}  [D_{ij}]^{\;\;\beta}_{\alpha}\Psi_{j\beta}, \quad \text{where} \\ [D_{ij}]^{\;\;\beta}_{\alpha} = \frac{1}{2}\delta_{ij} \delta^{\;\;\beta}_{\alpha} - K P_{ij} [H_{ij}]^{\;\;\beta}_{\alpha}.
    \end{split}
    \end{align}
    Here, $P_{ij} = 1$ if $i$ and $j$ are neighbours on the triangulation $T$, and $0$ otherwise. Note that the summation convention holds for spinor indices.
    Using Majorana relation \eqref{eq=majorana}, we can express the action using independent degrees of freedom
    
    \begin{equation}
     S_{T} =  \sum_{i, j}\Psi_{i\alpha}  [D_{ij}]^{\alpha\beta}\Psi_{j\beta} \quad \rightarrow \quad  [D_{ij}]^{\alpha\beta} = \frac{1}{2}\delta_{ij} \epsilon^{\alpha\beta} - K P_{ij}\epsilon^{\alpha\gamma} [H_{ij}]^{\;\;\beta}_{\gamma}
    \end{equation}Now, in order to make the indexing of this quantity clearer we repackage pairs of triangle and spinor indices into a single index, denoted by capital latin letters: $(i, \alpha) \rightarrow A := 2(i - 1) + \alpha$. Note that this map is $1$ to $1$ since $\alpha = 2 - (A\; (\text{mod}\; 2))$ and $i = 1 + \frac{A-\alpha}{2}$.
   Therefore, this induces a bijection for spinorial quantities:
    \begin{align}
    \begin{split}
    \Psi_{i\alpha} \rightarrow &\ \Psi_{A},
    \\
    [D_{ij}]^{\alpha\beta} \rightarrow &\ D_{AB},
    \end{split}
    \end{align}
    where $A$ and $B$ are integers from $\{1, 2, ... 2N\}$, where $N$ is the size of the triangulation. 
    The Dirac-Wilson operator is therefore represented as an antisymmetric $2N \times 2N$ matrix, and we are now ready to write the path integral of the theory as the sum over possible triangulations and Grassmann field configurations
       \begin{equation}
        Z = \sum_{T}Z_{T}=\sum_{T}\int\prod_{A} d \Psi_{A}e^{-\sum_{A, B}\Psi_{A}  D_{AB}\Psi_{B} }.
    \end{equation}
    Here $Z_{T}$ represents the fermionic path integral for a given triangulation $T$ and corresponds to the relative weight of the related geometry. This can be easily evaluated as the Pfaffian of the Dirac operator (ref)
    \begin{equation}\label{eq=pfaffian}
    Z_{T} = \text{Pf}\:D_{AB}(T) = \sqrt{\text{Det}D_{AB}(T)}
    \end{equation}Where we introduced the argument $T$ to make explicit the dependence of the triangulation. Note that in \eqref{eq=pfaffian} we implicitly used the positivity of the determinant, which was proved by Burda \ in \cite{burda1999wilson}.  
    In practice, when simulating the theory with MCMC, e.g.\ through the Metropolis\textendash Hastings algorithm \cite{buddMonte}, one computes the log of the determinant, since a direct computation of $D_{AB}$ is prone to cause floating-point overflow. This motivates the definition of a triangulation effective action
    \begin{equation}
    S_{T}^{\{eff\}} = - \ln{Z_{T}} = -\frac{1}{2} \text{logDet} D_{AB}(T).
    \end{equation}
    
\section{Hausdorff dimension}\label{hausdorff}
To test our method, we measured the scaling behaviour of the \textit{distance profile} $\rho_{T}(r)$, a well-studied geometric observable. 
In this section we will provide a brief description of its main properties, 
Let us first introduce a concept of distance on a triangulation: the \textit{graph distance} $d_{T}$ between two vertices $x$ and $y$ is defined to be the number of edges on the shortest path connecting $x$ and $y$. However, one should note that this form of distance converges, for large triangulations, to the geodesic distance, a more natural concept on smooth manifolds \cite{buddMonte}.
Now we can define the distance profile
\begin{equation}
\rho_{T}(r) \equiv \frac{1}{\mathcal{N}}\sum_{x,y}\mathbbm{1}_{(d_{T}(x,y) = r)}
\end{equation}
Where $\mathcal{N} = \frac{N+4}{2}$is the number of vertices in a triangulation of size $N$.
For an infinite system, the distance profile, averaged over all triangulations, has the scaling behaviour \cite{Budd:2022zry}
\begin{equation}
\lim_{N\to\infty}\mathbb{E}[\rho_{T}(r)] \propto r^{d_{W}}
\end{equation}where the critical exponent $d_{W}$ is called the \textit{Hausdorff dimension}, and it encodes the increase in connectivity with distance on a manifold.
For finite triangulations, the distance profile can only have support up to some maximum distance and therefore peaks at some value. This behaviour is characterized by the finite-size scaling relation \cite{ambjornflipmove, finitesizescaling}
\begin{equation}\label{eq=finitesizescaling}
\max_{r}\mathbb{E}[\rho_{T}(r)] \propto N^{1-\frac{1}{d_{W}}}.
\end{equation}
\section{Results}
We computed the Hausdorff dimension $d^{\text{spinor}}_{W} = 4.22 \pm 0.03$ for a system of free spinors using the methods outlined in the previous sections, by fitting $d_{W}$ in the formula \eqref{eq=finitesizescaling}, employing triangulations of sizes ranging from $80$ to $500$ triangles. 
Similarly, we also computed, the Hausdorff dimension of the critical Ising model, yielding the same result $d^{\text{Ising}}_{W} = 4.21 \pm 0.03$. This is to be expected since both systems have a central charge $c = \frac{1}{2}$.\\
Our values disagree with the previous estimates from Bogacz in \cite{bogacz2002spectrum}, i.e.\ $d^{B}_{W} = 2.87$. An important remark is that the Hausdorff dimension in \cite{bogacz2002spectrum} was obtained with a different method, i.e.\ by simulating an Ising model tuned to be dynamically equivalent to spinors. Therefore, our fit for the Ising model, in line with further supports the thesis.
Our measurements are consistent with the conjectural relation between the Hausdorff dimension and central charge proposed by Watabiki in \cite{watabiki1993analytic}, yielding $d_{W} \approx 4.2122$, and it is consistent with the bounds rigorously derived by Gwynne $4.1892 < d_{W} < 4.2156$\ \cite{gwynne2019bounds}, and with the state of the art numerical results from Budd, Ambjørn and Barkley \cite{ambjorn2013toroidal,barkley2019precision}.
As a practical note, we point out that computational complexity for the MCMC simulation of free fermions as we detailed scales as $N^{3}$, with $N$ the size of the triangulation. This is owed to the necessity to compute the determinant of the Dirac\textendash Wilson operator, and it poses a limit to its viability for lattices of size greater than $O(1000)$. For reference, in order to estimate the Hausdorff dimension of the free spinor model on lattices of size 500, we required approximately four hours of CPU time for a common laptop.

        \section{Outlook for non-Riemannian models}\label{interactions}
       To illustrate the greater flexibility of our fermion implementation, based on the Dirac-Wilson operator, we study in this section a slightly more elaborate interacting model, where a Dirac spinor $\psi$ is coupled with gravity through a torsionful connection for spinors $\:\omega_{a} = \frac{1}{2}(\Gamma_{a} + K_{a})\epsilon $, with contorsion components $K_{a}$, described by the following matter lagrangian
       \begin{equation}
       \mathcal{L} = \bar{\psi} \cancel{D}\psi + \frac{1}{2}\mu K_{a}K^{a} =  \bar{\psi} \gamma^{a} (e_{a}^{\nu}\partial_{\nu} + \frac{\Gamma_{a}}{2}\epsilon + \frac{K_{a}}{2}\epsilon) \psi + \frac{1}{2}\mu K_{a}K^{a}
       \end{equation}
       where $e_{a}^{\nu}$ is the \textit{zweibein} field, relating the vector coordinate basis to a local frame basis \cite{blagojevic2001gravitation}.
       This describes a Poincare gauge theory with non-dynamical torsion, as extensively discussed in Blagojevic's work \cite{blagojevic2001gravitation}. The equation of motion for the contorsion reads
       \begin{equation}
        \mu K^{a} = \frac{1}{2} \bar{\psi} \gamma^{a}\epsilon\psi.
       \end{equation}
       Since this equation provides an algebraic constraint for $K_{a}$, we expect that the theory
       \begin{equation}
           \mathcal{L'} = \mathcal{L} - g \bar{\psi} \gamma^{a}\epsilon\psi \: \bar{\psi} \gamma_{a}\epsilon\psi
       \end{equation}
       reproduces the dynamics of a free Dirac spinor, and therefore becomes conformal, when
       \begin{equation}
           g = g_{crit} \equiv \frac{1}{8\mu}.
       \end{equation}
       As a consequence, we could study critical $\mathcal{L'}$ theory on EDT with the following path integral associated to a given triangulation $T$
       \begin{equation}\label{eq=path_fourpoint}
           Z_{T} = \int\prod_{i}d\bar{\psi}d\psi e^{-S_{T}}.
       \end{equation}
    The action $S_{T}$ on EDT can be determined through a similar discretization scheme as the one we used for equation \eqref{eq=action}. This time, the action also contains a four-point interaction term for the spinors, so we can write it, schematically, as
    \begin{equation}
         S_{T} = S_{K} + \sum_{i, j}\bar{\psi}_{i} D_{ij} \psi_{j} -\sum_{i} g \bar{\psi}_{i} \gamma^{a}\epsilon\psi_{i} \: \bar{\psi}_{i} \gamma_{a}\epsilon\psi_{i}
    \end{equation}
    where we define $S_{K}$ to be the contribution to the action that only depends on the contorsion, and $D_{ij}$ is an analogous term to the Dirac\textendash Wilson operator in \eqref{eq=diracwilson}, this time also accounting for torsion effects.
    The path integral \eqref{eq=path_fourpoint} can then be computed as a series in powers of $g$ and simulated with MCMC in a similar way as with free fermions.
        \section{Conclusion}
         In this paper we proposed a new method for MCMC simulations of spinors on 2D euclidean dynamical triangulations, by computing the determinant of the Dirac-Wilson operator. This was tested numerically by fitting the Hausdorff dimension ${d^{\text{spinor}}_{W} = 4.22 \pm 0.03}$, which agrees with the current theoretical bounds from \cite{gwynne2019bounds} and disproves the previous measurements in \cite{bogacz2002spectrum}. Finally, we showed how our technique can be generalized to study interacting theories and non-Riemannian features of gravity on the lattice.

\section*{Acknowledgements}
  We are grateful for vital conversations with Zdzisław Burda and the useful correspondence with Timothy Budd, that provided key inputs for our work. 
  This work was performed using resources provided by the Cambridge Service for Data Driven Discovery (CSD3) operated by the University of Cambridge Research Computing Service (\href{www.csd3.cam.ac.uk}{www.csd3.cam.ac.uk}), provided by Dell EMC and Intel using Tier-2 funding from the Engineering and Physical Sciences Research Council (capital grant EP/T022159/1), DiRAC funding from the Science and Technology Facilities Council (\href{www.dirac.ac.uk}{www.dirac.ac.uk}), and ERC funding granted to Will Handley.

W.E.V.B. is grateful for the kind hospitality of Leiden University and the Lorentz Institute, and the support of Girton College, Cambridge.
    %
    
    	\bibliographystyle{unsrt}
    	\bibliography{main.bib}

	\end{document}